# Adaptive Energy Aware Data Aggregation Tree for Wireless Sensor Networks

Deepali Virmani[1], Tanu Sharma[2] & Ritu Sharma[3]

[1]Department of CSE, BPIT, GGSIPU, Delhi, India, 110085.
deepalivirmani@gmail.com

[2]Department of CSE, BPIT, GGSIPU, Delhi, India, 110085.
tanusharma1217@gmail.com

[3]Department of CSE, BPIT, GGSIPU, Delhi, India, 110085.
ritusharma1108@yahoo.com

*Abstract*

*To meet the demands of wireless sensor networks (WSNs) where data are usually aggregated at a single source prior to transmitting to any distant user, there is a need to establish a tree structure inside to aggregate data. In this paper, an adaptive energy aware data aggregation tree (AEDT) is proposed. The proposed tree uses the maximum energy available node as the data aggregator node. The tree incorporates sleep and awake technology where the communicating node and the parent node are only in awake state rest all the nodes go to sleep state saving the network energy and enhancing the network lifetime. When the traffic load crosses the threshold value, then the packets are accepted adaptively according to the communication capacity of the parent node. The proposed tree maintains a memory table which stores the value of each selected path. Path selection is based on shortest path algorithm where the node with highest available energy is always selected as forwarding node. By simulation results, we show that our proposed tree enhances network lifetime minimizes energy consumption and achieves good delivery ratio with reduced delay.*

## 1. Introduction

Wireless sensor network (WSN) is a collection of densely deployed sensor nodes for the collection and dissemination of environmental data. These Wireless sensor networks have wide range of application field from Environmental monitoring like Forest fire detection, air pollution monitoring and green house monitoring to Industrial monitoring and are also used greatly for the military purposes[1]. WSN accurately monitor and control the physical environment from the remote locations as well. Sensor nodes used in these WSN are limited in energy, power, memory and computational capacity, so it has become necessary to design the deployment techniques in such a manner so that the available energy, power and memory can be used in an efficient manner [2].

Data Aggregation is used in these WSN to maximize the life time of the network by gathering and aggregating the data in an energy efficient manner and by reducing the medium access layer contention. Data aggregated is further transferred to some sink node by following some path [10]. Actually data is transmitted from source node to the sink node in multi hop manner to the neighboring nodes which are further nearer to the sink node. This transmission of data from source to sink in multi hop manner using neighboring nodes should be efficient in such a manner that energy and power consumption is less as compared to direct transmission form source to sink. Data aggregation can be further of two types, Lossless [8] and Lossy aggregation [9]. Lossy aggregation is used in the situation when the load of data to be transmitted exceeded the system capacity. And Lossless aggregation is used when the load is under the threshold limit or when there is comparative less data has to be transferred.

In this paper, we have proposed an energy aware data aggregation tree which incorporates sleep and awake technology periodically in combination with other energy efficient techniques to provide us with better results. Section 2 of this paper discusses the literature survey in field of energy efficient data aggregation. In section 3, we have explained the basic idea behind our tree followed by complete tree formation. Section 4 contains the simulation results. Section 5 discusses conclusion and future scope.

## 2. Literature survey

For applying compressive theory to the data aggregated in WSN, the first design has been presented by Chong Luo et a [3]. This technique is able to reduce global scale communication cost without increasing the complications in transmission control. This scheme can also handle the abnormal sensor readings. They have used ns-2 [7] simulator to validate the analysis of the network capacity of the proposed compressive data aggregation tree. A semi structured approach using dynamic forwarding on an implicitly constructed structure composed of multiple shortest path first trees to enhance network stability, a tree on DAG (ToD) has been proposed by Prakash G el at[4]. The basic principle in ToD is that adjacent nodes in the graph will have low stretch in one of these ToD trees and leading to early aggregation of data. They concluded that efficient data aggregation in the network can be achieved by their semi-structured approach, using simulations based on 2,000-node Mica2 based network.

Qiang Ma et al [5] have presented the general architecture for seamlessly integrating data aggregation into WSN and for presenting one error correction algorithm for the data algorithm by exploiting the inherent correlations existing between sensor nodes. Structure free data aggregation for event based sensor network has been explored potentially by Kai Wei Fan et al [6]. Data Aware Any cast at MAC layer and randomized waiting at application layer- the corresponding mechanisms are proposed. Simulation results have been used to show the performance and potential of structure free data aggregation.

## 3. Adaptive energy aware data aggregation tree

We propose an adaptive energy aware data aggregation tree which incorporates sleep and awake technology periodically. Our proposed tree uses maximum available energy node as parent node for data aggregation.

Our work differs from the previous work in the following manner. Our tree has following features.

- It uses the node with maximum Available energy as the parent node.
- It focuses on the sleep and awake technique thus minimizing the energy loss.
- It focuses on the communication capacities of the nodes.
- Memory table for the path used will be maintained for the future use.

In our Adaptive Energy Aware Data Aggregation Tree, we have used the shortest path first algorithm with respect to Available Energy (Eavail) for transmitting data from source to the parent node. None of the previous works have collaborated SPF with Eavail.Our proposed tree is refreshed after every t seconds where t is the real time delay and after every t seconds a new parent (if required) is formed.

### 3.1 Parent formation

This is the first step of the proposed tree aggregator tree where the parent node for aggregation is selected. Every node in the network broadcasts its available energy (Eavail). The node with the highest Eavail becomes the parent node of that network. Consider a network as shown in figure 1 having nodes A, B, C, D, E, and F as the communicating nodes with their respective available energy (Eavail) and communication capacity.

Every node broadcasts its Eavail and then the node with the maximum Eavail becomes the parent node (PA) of that network (in the above scenario, node A is the parent node with the maximum Eavail shown in figure2).

After becoming the parent node, A node broadcasts its status as 'awake'. As soon as any node changes its status to 'awake', rest all other nodes automatically go to 'sleep' state. Only the parent and communicating node remains in active state. Rest all other nodes remain in passive state. This helps in minimizing the energy consumption and enhancing the network lifetime.

Eavail is calculated by

$$E_{avail} = E_b(t2) - E_b(t1) + \int_{t1}^{t2} P_c(t) \qquad (1)$$

Where,

Pc -- power consumption of the network

Eb(t1) -- battery level of node at time t1

Eb(t2) -- battery level of node at time t2

$$P_c = O(\frac{Pt}{d^{\alpha}}) \qquad (2)$$

Where,

α -- 2-4

Pt -- transmitted power

d -- Distance

Pc – power consumption

Energy consumed by node is calculated as follows

$$E_{con} = \frac{Vin}{R} \int_{t0}^{t1} Vr(t) dt \qquad (3)$$

Where,

Vr(t) --voltage across test resistance

Vin --input voltage

Energy consumed by network is estimated by the size of the network: m * size + b

Where,

m and b -- linear coefficients

But in the case when more than one node is having same Eavail then, the node with more communication capacity will become the parent node of the network.

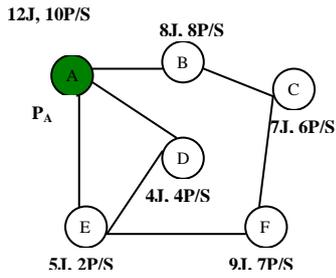
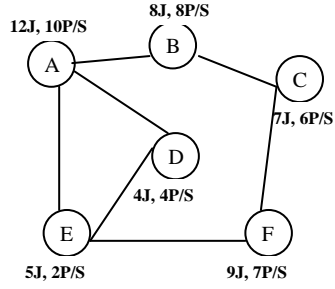

Figure1. Scenario of WSN           Figure2. Node A selected as parent

Notation for Our Proposed Parent Formation Algorithm
Ni, Nj = ith node where, i ε 1 to n
n = number of nodes of network
Ei = available energy of the ith node
Emax = maximum energy
Pnode = parent node
Nmax = node with maximum energy
CC = communication capacity

---

Algorithm for Parent Formation

1. Emax = 0
2. for every Ni, where i ε 1 to n
3. Broadcast available energy Ei
4. End for
5. for each Ni, i ε 1 to n
6. Compare Ei and Emax
    i. If Ei>Emax
7. Ei = Emax
8. Nmax = Ni
    i. If Nmax = Ni = Nj
    ii. Compare CCNi and CCNj
    iii. If CCNi>CCNj
    iv. Ni broadcasts status to the network as Pnode
    v. Ni = Pnode
    vi. Else
9. Nj broadcasts status to the network as Pnode
10. Nj = Pnode
11. End for
12. Pnode→awake
13. Rest nodes → sleep

---

### 3.2 Communication capacity

Communication capacity defines the data handling capacity of the nodes in the real world. Figure 3 explains the concept of Communication capacity. In the network given below, node A is the parent node (PA) with maximum Eavail and the communication capacity of 10packets/sec. As PA is the parent node with communication capacity of 10 packets, it will aggregate data coming from the other nodes. Suppose node 'C' wants to send data to node 'A'. So after receiving 6 packets from node 'C', node 'A' is left with the capacity of 4 more packets to be aggregated(as its communication capacity is 10 packets). Therefore, when node 'F' will send 7 packets to PA, it will be asked either to prioritize its packets and to send only 4packets or to wait for the network to get refreshed.

Real time Communication Capacity

$$\text{RTCC} = B\sum_{x} Ut_x(t) \quad (4)$$

Where,

$$\sum Utx(t) = \sum_{P_i \,\varepsilon\, Kx(t)} \frac{T_i}{D_i} \quad (5)$$

Where,

Utx(t) = utilization factor

Di is Distance of each node with respect to sink node

Ti is Transmission Time with each packet

B is Bandwidth utilization

Pi ratio of packet size and effective bandwidth

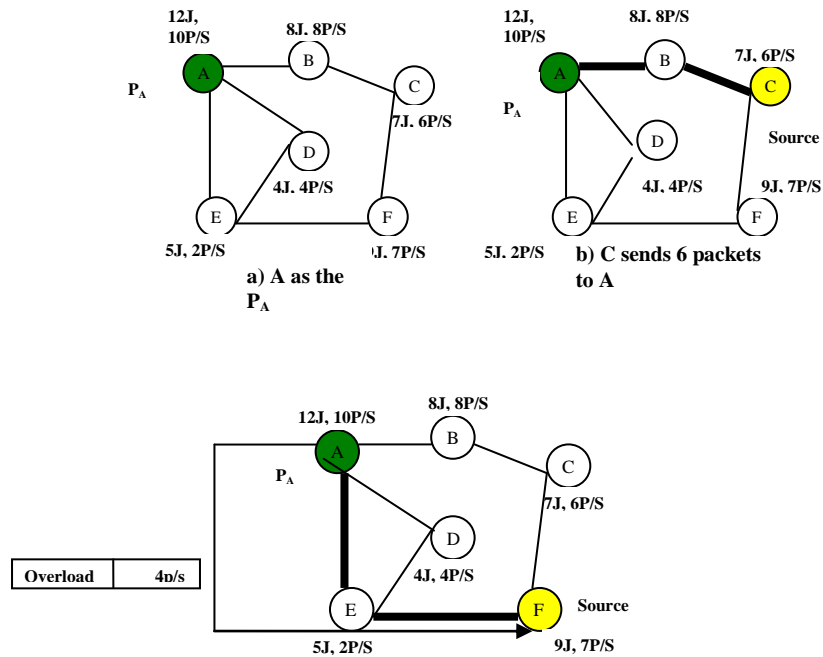

a) A as the P_A

b) C sends 6 packets to A

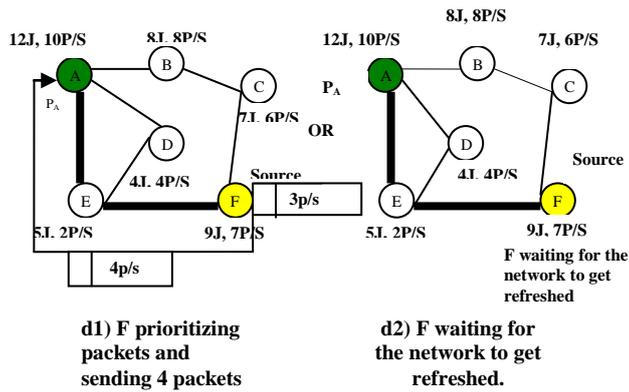

d1) F prioritizing packets and sending 4 packets

d2) F waiting for the network to get refreshed.

c) F sends 7 packets but A left with 4 packets capacity (as C has already sent 6 packets)

Figure 3 Communication Capacity

### 3.3 Sleep and awake mode

- The nodes of the network will either be in a sleep mode or awake mode.
- Parent node of the network (Pnode) will always be in an awake mode.
- The Pnode will remain awake till the network gets refreshed after 't'seconds, where 't' depends on the delay that can be supported by the real time scenario.
- After every cycle, the new Pnode will broadcasts its status as awake rest of the nodes will remain in sleep mode.
- Whenever a node wants to transmits a message, it changes its status to awake.
- The intermediate nodes remain awake only for the time they are transferring data from the source node to the sink node. Rest they always remain in a sleep mode.
- The energy of the Pnode will decrease by unit 1 i.e $Ei = Ei - 1$ ( where Ei is the energy of the Pnode) and energy of the rest of the nodes will decrease by α which is directly proportional to the time period for which the node is in awake mode in order to transmit data.

### 3.4 Path selection

Once a parent node is selected and some other node in the network wants to transfer some data to the parent node, then it becomes necessary to select some path from transmitting node to the parent node. This path selection should be efficient in such a manner that it saves energy of the nodes thereby increasing the lifetime of the sensor network. For this purpose we have proposed the Shortest Path first algorithm with respect to the available energy (Eavail). It means that when the transmitting node has to send data to the parent node it will choose its nearest neighbor with maximum Eavail (available energy) as the next recipient in the path. This same process will continue in multi hop manner until the parent node is reached. Moreover our proposed tree maintains a database for the paths been chosen. This table ( figure 4) will contain the entries for the path from transmitting node to the parent node. Whenever some node requires to send data to parent node, first the database is scanned if there is already an entry in the database the same path will be used for communication. Else an alternative path is selected on the basis of the proposal. This will minimize the time in searching a path therefore minimize delay but the same path may lead to nodes with less available energy. This can be the drawback of our proposed work.

The example given below describes the proposed method.

Now F will always use the path F – C – B – A that has been stored in the memory table. Similarly all the other nodes will store their respective paths (considering A as the parent node) in memory table for future reference by following the same technique. The memory table will look like:

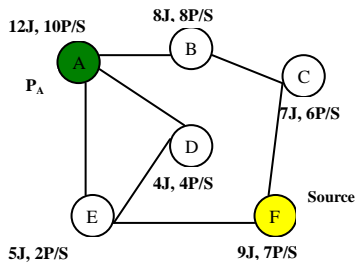

Figure 4 Path Selection

1) Here, F sends data to A by following the path using SPF technique F – C – B – A.
2) Once followed, F will store the path in memory table and will always refer it for future use.

| PARENT NODE | TRANSMITTING NODE | PATH |
|---|---|---|
| A | F | F – C – B - A |

| PARENT NODE | TRANSMITTING NODE | PATH |
|---|---|---|
| A | F | F – C – B - A |
| A | C | C – F – E – A |
| A | D | D – E – A |
| A | B | B – C – F – E - A |
| A | E | E – F – C – B - A |

Similarly nodes will store their respective paths for all the parent nodes in the network by using the technique of shortest path first algorithm using the concept of maximum energy available ( Eavail).

Path Selection Algorithm

Path Select( )
1. TnodeBroadcast Request.
2. Neighbour Unicast Eavail.
3. Establish Connection {Connection → Max Eavail(Nnode), where Nnode is one hop neighbour
4. Repeat step 3, till Pnodeis reached.
5. Path: New Path.
6. Update MTable.

**3.5 Adaptive energy aware data aggregation tree formation (AEDT)**

In our proposed tree, the whole network will be refreshed after every t seconds where t is the delay time depending on the real time scenario. Whenever some node will require to send some data, it will look for the parent node at that time. Then the transmitting node will search for some existing path in the memory. If no corresponding path is found it will follow the path formation algorithm described in the section 3.4. The parent node on receiving the data will either send under load or overload message depending on its communication capacity. Under load message will act as acknowledgement for the transmitting node. Over load message will tell the transmitting node that the size of data packets is too more than the communication capacity of the parent node. The transmitting node will either wait for the time period till the network is refreshed and a new parent is formed or it will prioritize the data packet and will send data in chunks. Meanwhile a special care will be taken about the sleep and awake mode of nodes. Only parent node will remain awake for the t seconds. When they are sending data or passing data respectively as discussed in the section 3.3. By using the concept of sleep and awake we have further reduced the energy consumption and thus making our proposed tree more energy efficient. Transmitting node and the intermediate nodes will remain awake only for the time period

Steps for the formation of AEDT

Notations used in the proposed algorithm
Pnode= Parent Node
Tnode= Transmittting Node
Inode= Intermediate Node
Link(A,B)= Connection from Node A to Node B.
Mtable= Memory table
DT = Number of packets to be sent by Transmitting Node.

Various functions defined in the formation of proposed tree
SEARCH( ): Broadcast and search for Parent Node.
CHECK( ): Checks the memory table for the desired path.
PATH_SELECT( ): Described in Section 3.4
PRIORITIZE( ):Described in Section 3.4

---

ALGORITHM FOR AEDT

1. Refresh network
2. Call PARENT_FORM()
3. Parent = Pnode
4. Tnode SEARCH Pnode
5. Tnode CHECK Mtable
6. Initialize Path:
7. PATH: LINK(Tnode,Pnode)
8. If PATH exists
9. T→ PATH
10. Else
11. Call PATH_SELECT()
12. Refresh Mtable.
13. If DT <CCPnode
14. Message: Accept and Acknowledge
15. Else
16. Message: Overload
17. If Message: Overload, then
18. Tnode→ WAIT(t)
19. OR
20. Tnode→ Prioritize()
21. Eavail(Pnode) →Eavail(Pnode) -1
22. Eavail(Tnode)→Eavail(Tnode) -1
23. Eavail(Inode) →Eavail(Inode) – 1
24. Refresh Network.

# 4. Simulation Results

## 41. Simulation Setup

The performance of our AEDT technique is evaluated through NS2 [21] simulation. A random network deployed in an area of 500 X 500 m is considered. We vary the number of nodes as 20, 40….100. Initially the nodes are placed randomly in the specified area. The base station is assumed to be situated 100 meters away from the above specified area. The initial energy of all the nodes assumed as 3.1 joules. In the simulation, the channel capacity of mobile hosts is set to the same value: 2 Mbps. The distributed coordination function (DCF) of IEEE 802.11 is used for wireless LANs as the MAC layer protocol. The simulated traffic is CBR with UDP source and sink.

## 4.2 Performance Metrics

The performance of AEDT is compared with the CDG [10] and ToD [12] protocols. The performance is evaluated mainly, according to the following metrics.

1. Average end-to-end Delay: The end-to-end-delay is averaged over all surviving data packets from the sources to the destinations.
2. Average Packet Delivery Ratio: It is the ratio of the number .of packets received successfully and the total number of packets transmitted.
3. Energy Consumption: It is the average energy consumption of all nodes in sending, receiving and forward operations.
4. Network Lifetime: The network lifetime is the time till the first node runs out of energy.
   The simulation results are presented in the next section.

## 4.3 Simulation Results

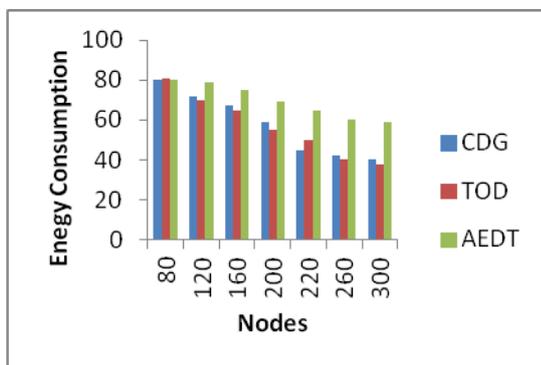
Figure5. Energy Consumption

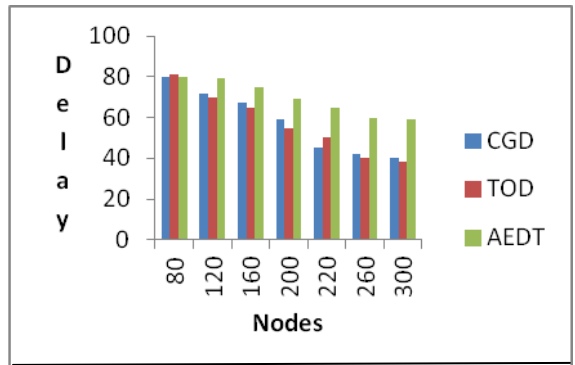
Figure 6. Average End to End Delay

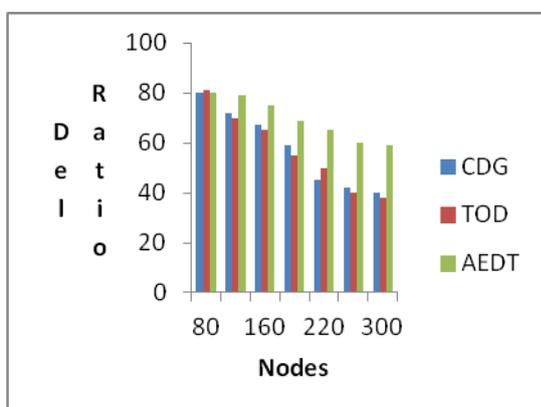
Figure7. Delay Ratio

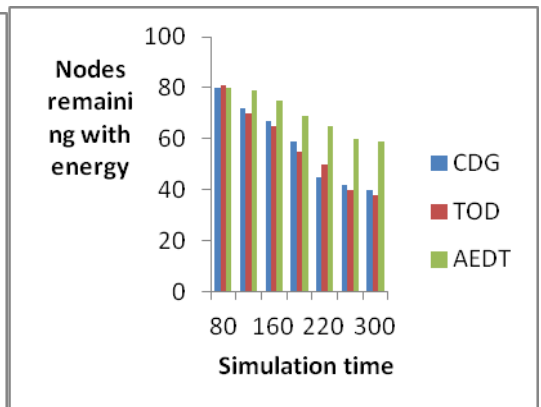
Figure 8. Nodes Remaining with Energy

As shown as in figure 5 AEDT technique has less energy consumption when compared with ToD and CDG, since it has the energy efficient tree. The average end-to-end delay of the proposed AEDT technique is less as compared with ToD and CDG. Figure 6 shows end to end delay. Due to adaptive nature delivery ratio of proposed tree is high as compared to existing ToD and CDG (figure 7). As shown in figure 8 network lifetime is enhanced with AEDT as compared to ToD and CDG because highest energy available node is always used for communication.

## 5. Conclusions

In this paper, we have proposed an adaptive energy aware aggregation tree for wireless sensor networks. In the proposed tree node with highest energy available is selected as the parent node for aggregation. The proposed tree incorporates sleep wait technique where only the parent node and the communicating node are in awake state rest all the nodes go to sleep state, where in the intermediate nodes periodically go to awake state if they have any message to be forwarded. Shortest distance path with maximum available energy is selected for communication. Once a path is selected it is stored in memory for further use. Each node estimates its traffic load when it receives the data from the sources. The parent node estimates the total traffic load in the system and sends an OVERLOAD packet to the sources if it is greater than its communication capacity. As soon as the source node receives the OVERLOAD packet it decides whether to send message in chunks depending on the communication capacity of parent node or waits till the refresh time. If the traffic load is less than the communication capacity the parent node accepts the message for further communication. The proposed tree provides a reliable transmission environment by minimizing energy consumption. Simulation results prove that our proposed technique achieves good delivery ratio with reduced delay and minimized energy consumption. As well as the proposed tree enhances network lifetime by always using maximum energy available node for transmission.

## Authors

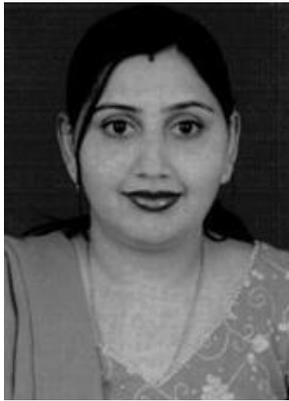

**Deepali Virmani** Received the B.E. Degree from Maharishi Dayanand University, Rohtak (Haryana) in 2001 & M.Tech Degree From Guru Gobind Singh Indraprastha University, Delhi in 2005. Ph.D Thesis submitted in field of sensor networks. She is working as an Associate Professor in Dept of CSE in BPIT, affiliated with GGSIPU, Delhi. Her Research Interest includes wireless sensor network and adhoc network. She is a member of IEEE, IETE & ISTE India. She is the member of editorial board of various international journals like International Journal of Advanced Engineering & Applications International Journal of Network Security International Journal of Engineering and Advanced Technology (IJEAT) International Journal of Soft Computing and Engineering (IJSCE).

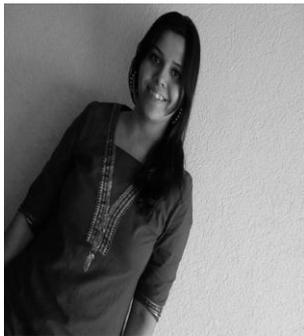

**Tanu Sharma** Received the B.E. Degree from Maharishi Dayanand University, Haryana in 2009 & pursuing M.Tech Degree from Delhi Technical University, Delhi. She is working as an Assistant Professor in Dept of CSE in BPIT, affiliated with GGSIPU, Delhi. Her Research Interest includes wireless sensor network and adhoc network.

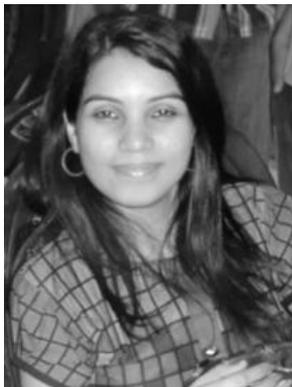

**Ritu Sharma** Received the B.Tech. Degree from Guru Gobind Singh Indraprastha University, Delhi in 2008 & pursuing M.Tech Degree from Mahamaya Technical University (former UPTU), Noida. She is working as an Assistant Professor in Dept of CSE in BPIT, affiliated with GGSIPU, Delhi. Her Research Interest includes wireless sensor network and adhoc network.